\def\GraphIn#1#2#3#4#5#6{
  \dimen0=#2in%
  \hbox to #4\dimen0{%
   \dimen0=#3in%
   \vbox to #4\dimen0{
    \vss
      \special{ps: plotfile #1 asis}
      \special{ps::
         DissplaCheckPoint restore
      }
    }%
    \hss
   }
 }

\def\IC{\relax\hbox{$\inbar\kern-.3em{\rm C}$}}
\def\IR{\relax{\rm I\kern-.18em R}}
\font\cmss=cmss10 \font\cmsss=cmss10 at 7pt
\def\IZ{\relax\ifmmode\mathchoice
{\hbox{\cmss Z\kern-.4em Z}}{\hbox{\cmss Z\kern-.4em Z}}
{\lower.9pt\hbox{\cmsss Z\kern-.4em Z}}
{\lower1.2pt\hbox{\cmsss Z\kern-.4em Z}}\else{\cmss Z\kern-.4em Z}\fi}
 
\input harvmac
 
\Title{\vbox{\baselineskip12pt\hbox{UICHEP-TH/94-3}}}
{\vbox{\centerline{Nonabelian Vortices on Surfaces and Their
Statistics}}}
 
\centerline{Lee Brekke, Sterrett J. Collins\footnote{$^*$}{lobo@uic.edu}, 
and Tom D. Imbo\footnote{$^{\dagger}$}{imbo@uic.edu}}
 
\bigskip\centerline{Department of Physics}
\centerline{University of Illinois at Chicago}
\centerline{845 W. Taylor St.}
\centerline{Chicago, IL \ 60607-7059}

\vskip 1.0in
 
We discuss the physics of topological vortices moving on an arbitrary surface 
$M$ in a Yang-Mills-Higgs theory in which the gauge group $G$ breaks to a 
finite subgroup $H$. We concentrate on the case where $M$ is compact and/or 
nonorientable. Interesting new features arise which have no analog on
the plane. The consequences for the quantum statistics of vortices are 
discussed, particularly when $H$ is nonabelian.

\Date{}
 
\def\bra#1{\left\langle #1 \right|}
\def\ket#1{\left| #1 \right\rangle}

\newsec{Introduction}

The notion of indistinguishability lies at the core of many quantum phenomena.
For example, amplitudes contributing to a given process that correspond to 
distinct classical possibilities will interfere if and only if the experimental 
setup cannot distinguish between them. As a result, when two particles which 
are indistinguishable in a quantum system scatter, their direct and exchange 
amplitudes interfere. Studying the implications of this basic fact for large 
numbers of indistinguishable particles is the subject of quantum statistical 
mechanics. We stress that indistinguishability is not a fundamental property of 
particles but depends, in general, on the details of the relevant experiment. 
A hypothetical particle which is identical to the electron in all respects 
except for an extremely small mass difference well beyond current experimental 
resolution would still display the usual interference phenomena with ``real'' 
electrons. These new particles are {\it indistinguishable} from electrons even 
though they are not {\it identical} to them \ref\rus{V.~L.~Lyuboshitz and 
M.~I.~Podgoretskii, Sov. Phys. JETP {\bf 28} (1969) 469\semi R.~Mirman, Nuovo 
Cim. {\bf 18B} (1973) 110.}. The interference would disappear with the advent 
of a more precise measuring apparatus capable of detecting the ``hidden'' mass 
difference. The bottom line is that while identical particles are indeed 
indistinguishable, the converse is not necessarily true.

The hidden difference between distinct yet indistinguishable particles need 
not be in some continuous parameter like mass. They may possess distinct
values of a {\it discrete} label which cannot be detected in the experiment at 
hand. For instance, in an experimental setup which does not couple to spin, 
spin up and spin down electrons are indistinguishable and seem to obey 
parastatistics of order two \ref\para{H.~S.~Green, Phys. Rev. {\bf 90} (1954)
270.}. (More generally, particles with spin $s$ will behave as paraparticles 
of order $2s+1$.) If we could ``turn off'' all interactions in the universe 
which couple to spin, then different spin states would be {\it in principle} 
indistinguishable (although still not identical), and the particles would obey 
parastatistics at the fundamental level. We can play the same game with 
color, isospin and numerous other internal degrees freedom. (Indeed, the 
parastatistical treatment of quarks \ref\green{O.~W.~Greenberg, Phys. Rev. 
Lett. {\bf 13} (1964) 598.} predates the introduction of color 
\ref\han{M.~Y.~Han and Y.~Nambu, Phys. Rev. {\bf 139} (1965) B1006.}.)
 
In two spatial dimensions, it is possible to introduce internal labels which
lead to statistical behavior other than parastatistics. These labels, however, 
must be coupled to the spatial degrees of freedom in a rather intricate way.
A beautiful class of models containing particle-like objects possessing such 
structure is that of nonabelian vortex systems. More specifically, consider a
(2+1)-dimensional spontaneously broken gauge theory in which the (simply
connected) gauge group $G$ is broken down to a {\it finite} subgroup $H$.
Such a theory contains topologically stable vortices labelled by $h\in H$.
A vortex configuration specified by $h_1$ and one labelled by any conjugate
$h_2=h^{-1}h_1h$ are indistinguishable, even though they are not identical
unless $h$ commutes with $h_1$ \ref\bdfi{L.~Brekke, H.~Dykstra, A.~Falk and 
T.~D.~Imbo, Phys. Lett. {\bf B 304} (1993) 127.}\ref\lp{H.-K.~Lo and
J.~Preskill, Phys. Rev. D {\bf 48} (1993) 4821.}. (The precise meaning of 
indistinguishability used here will be given below.) This result is 
responsible for the exotic statistical behavior of vortices in many examples. 
In particular, it generically leads to the existence of {\it nonabelian} 
statistics.

In this paper we will show that new features arise if the surface on which the 
vortices are moving is nonorientable and/or compact. (The compact case has 
already been discussed in some detail --- although with a different emphasis 
--- in \ref\kml{K.-M.~Lee, Phys. Rev. D {\bf 49} (1994) 2030.}.) For example, 
on nonorientable surfaces a vortex specified by the ``flux element'' $h$ and 
one labelled by its inverse $h^{-1}$ are also indistinguishable. This shows 
that spatial topology plays a role in determining what we can consider to be 
indistinguishable. On a compact surface $M$ there is another interesting 
feature. Namely, we must identify any two $n$-vortex states on $M$ which differ 
by a global $H$ gauge transformation. One way of seeing this is that the action 
of the braid group $B_n(M)$ on the $n$-vortex states is not well-defined unless 
this identification is made. (This fact was overlooked in \kml , although the 
physics results in that paper are still valid.) Both of these results have
consequences for the statistics that a system of vortices may obey, as we will 
see in detail below.

\newsec{Nonabelian Vortices on the Plane and Their Statistics}

\subsec{Two Notions of Indistinguishability}

We begin with a few preliminaries concerning vortex systems on the plane 
$\IR^2$. Many of these results will generalize to vortices on an arbitrary 
surface. We will restrict our attention to configurations in which the gauge 
and Higgs fields take on vacuum values everywhere in space, except in the 
cores (assumed to be ``small'') of a finite number $n$ of isolated vortices. 
For $n=1$, the flux of the vortex is defined as the path-ordered exponential 
(or {\it holonomy}) of the gauge field around a loop $C$, based at some fixed 
point $x_0$, and circling the vortex once (avoiding its core) with a given 
orientation. Each such exponential yields an element of the unbroken subgroup 
$H=H(x_0)$, which is the stabilizer of the Higgs field at $x_0$. This flux is 
invariant under continuous deformations of $C$ which do not cross the vortex 
and do not move the basepoint $x_0$. So we see that for $n=1$, any two choices 
of $C$ which have the same orientation will yield the same flux for the vortex. 
(Changing the orientation just changes the flux to its inverse in $H$.) More 
generally, when determining the flux of a vortex occupying the position $x_i$ 
in the presence of one or more spectator vortices, one wants a loop $C_i$ 
(still based at $x_0$) which avoids all vortex cores and encloses only the 
vortex at $x_i$. But in contrast to the situation for $n=1$, there are now 
various topologically distinct choices which meet these criteria for any 
fixed orientation. That is, there exist acceptable, similarly oriented choices 
for $C_i$ which cannot be deformed into each other without crossing at least 
one vortex in the process. Moreover, the flux of the vortex at $x_i$ will, in 
general, depend on this choice if $H$ is nonabelian \ref\buch{M.~Bucher, Nucl. 
Phys. {\bf 350} (1991) 163.}. One simply has to pick a specific $C_i$ and stick 
with it. Another subtlety arises since a vortex (with flux $h$) may circle the 
basepoint $x_0$ --- a process that causes {\it all} vortex fluxes to be 
conjugated by $h$ \buch . (Again, this can  only be nontrivial if $H$ is 
nonabelian and $n\geq 2$.) Since the choice of a basepoint is completely 
arbitrary, this conjugation is unphysical and merely complicates the 
description of the system. By moving $x_0$ out to spatial infinity, this 
problem can be avoided. Note also that vortex fluxes are not, in general, 
invariant under a gauge transformation $g(x)$. Instead, all the fluxes in a 
given configuration become conjugated by $g(x_0)$ (which is assumed to be in 
$H(x_0)$ so that the Higgs field at $x_0$ is unchanged). However this is not a 
problem, since we can break up every gauge transformation $g(x)$ into a 
product of a {\it global} gauge transformation by $g_0=g(x_0)$, followed by a 
{\it local} transformation ${\hat g}(x)$ satisfying ${\hat g}(x_0)=e$. We need 
only require invariance of physical quantities under these latter 
transformations to obtain a sensible theory. The conjugation of fluxes under 
global $H$ gauge transformations simply amounts to a relabelling of the 
vortices.

We now come to an important point (see \buch\ref\bais{F.~A.~Bais, Nucl. Phys. 
{\bf B170} (1980) 32\semi L.~Krauss and J.~Preskill, Nucl. Phys. {\bf B341}
(1990) 50.}\ref\wz{F.~Wilczek and Y.-S.~Wu, Phys. Rev. Lett. {\bf 65} (1990) 
13.}). Consider taking a single vortex in an $n$-vortex configuration for an 
excursion in the plane which avoids all other vortices and finally returns the 
chosen vortex to its original position. Assume that the initial flux of this 
vortex is $h$ as measured with respect to some fixed loop $C$ based at a point 
$x_0$ at infinity. At the end of this trip, the flux of the vortex will no 
longer be $h$ in general, but will have changed to some conjugate of $h$. This 
is true even though the vortex did not encounter any forces throughout the 
process; it only moved through regions of vacuum. The specific final flux 
(which is still measured with respect to the loop $C$) will depend on the path 
taken by the vortex, as well as the fluxes of the other spectator vortices in 
the system, as we shall see more precisely later. In a similar manner, we may 
consider moving a {\it pair} of vortices around in the plane. Assume that one 
is located at $x_1$ and has flux $h_1$ (measured with respect to a loop $C_1$),
while the other is at $x_2$ and has flux $h_2$ (measured with respect to a 
loop $C_2$). After traversing paths which in general may be very complicated 
--- winding around other spectator vortices as well as each other --- the 
vortex initially at $x_1$ ends up at $x_2$ while that starting at $x_2$ 
finishes at $x_1$. The final flux of the vortex which ends up at $x_1$ 
(measured using $C_1$) will be a conjugate of $h_2$, while the final flux of 
the vortex ending up at $x_2$ (measured using $C_2$) will be a conjugate of 
$h_1$. (The specific values depend on the details of the paths.)

We can now use the above discussion to define the notion of vortex 
indistinguishability. Actually we will define both a {\it weak} and a {\it 
strong} version of indistinguishability. In the strong version, we consider a
configuration containing $n$ vortices located at the positions $x_i$ with
respective fluxes $h_i$ measured using loops $C_i$, $i=1,\dots ,n$. The 
$h_j$ and $h_k$ vortices are said to be indistinguishable {\it in this 
configuration} if there exists an $n$-vortex path (or {\it braid}) $(x_1(t),
\dots ,x_n(t))$, $0\leq t\leq 1$, satisfying $x_i(0)=x_i$ for all $i$, $x_j(1)
=x_k$, $x_k(1)=x_j$, $x_i(1)=x_i$ for $i\neq j,k$, and such that the flux of 
the vortex occupying the position $x_i$ at $t=1$ is still $h_i$ for all $i$. 
That is, each of the vortices returns to its original position except the ones
originally at $x_j$ and $x_k$ which exchange positions. Moreover, the final 
configuration is {\it identical} to the original one. This definition assures 
us that the direct and exchange contributions to the scattering of the two 
indistinguishable vortices in this background will interfere coherently \lp . 
Note that by the discussion at the end of the previous paragraph, the final 
flux at $x_j$ will be a conjugate of $h_k$ while the final flux at $x_k$ is a 
conjugate of $h_j$. Thus, a necessary condition for two vortices to be 
strongly indistinguishable is that their corresponding fluxes are conjugate. 
However, whether or not two vortices with specific conjugate fluxes are indeed 
indistinguishable in a given configuration will depend on the nature of the 
other fluxes around. Thus, this version of indistinguishability is 
{\it background-dependent}.

The definition of weak indistinguishability is that two vortices with fluxes 
$h_1$ and $h_2$ are indistinguishable if there exists {\it some} configuration 
containing them in which they are indistinguishable in the strong sense. As a 
result we have that two vortices on the plane (or indeed any orientable 
surface) are weakly indistinguishable if and only if their fluxes are 
conjugate. That is, for conjugate vortices there always exists some background
in which they are strongly indistinguishable. (This follows straightforwardly
from the results of Section 2.2 below.) It is certainly true that two 
conjugate vortices do not {\it have} to be treated as indistinguishable in a 
configuration in which there is no exchange braid as above. However, they still 
{\it can} be treated as indistinguishable with no inconsistency arising. This 
is analogous to the situation for spin up and spin down electrons in a system 
with no magnetic fields, or the proton and the neutron in the absence of 
isospin-violating processes --- you can treat them as either distinguishable 
with no internal degrees of freedom, or indistinguishable with internal degrees 
of freedom. Its your choice --- the resulting amplitudes remain the same. The 
analog of spin or isospin flip for the two conjugate vortices is a global 
gauge transformation by the element of $H$ which conjugates one vortex flux 
into the other. It is also nice that with this definition we do not to have to 
worry about the background in order to decide whether or not to treat two 
vortices as indistinguishable.

Another motivation for the weak version of indistinguishability comes from
the realization that the discussion so far has ignored the possibility of the
creation and annihilation of virtual vortex-antivortex pairs from the vacuum
\bdfi . Thus, in some sense, {\it every} configuration contains a vortex of 
every flux, even if only for a very short time. As a consequence of this
consider, for example, the physical process of measuring the flux $h$ of a
vortex by an ``Aharonov-Bohm'' scattering experiment utilizing particles 
carrying $H$-charge \ref\acm{M.~Alford, S.~Coleman and J.~March-Russell, Nucl. 
Phys. {\bf B351} (1991) 735.}. More precisely, consider a beam of particles 
each of which is in a state $\ket{u}$ (in charge space) which transforms 
according to some faithful representation $D$ of $H$. Initially this beam is 
heading toward the vortex of interest, but before it gets there it is split 
into two pieces, one of which goes to the left of the vortex and the other to 
the right. The two beams are then recombined on the far side of the vortex. At 
some observation point located beyond the point of recombination, the particles 
in the beam will be in the state $\psi_1\ket{u}+\psi_2D(h)\ket{u}$, where 
$\psi_1$ and $\psi_2$ are complex functions of position which can be determined 
in a straightforward manner. Thus, by measuring the probability density at the 
observation point, we can determine the matrix element $\bra{u}D(h)\ket{u}$. 
By performing a sufficient number of such experiments with charged particles 
initially in the internal symmetry state $\ket{u}+\alpha\ket{v}$, $\alpha$ a 
complex number, it seems as though all of the matrix elements $\bra{v}D(h)
\ket{u}$, and hence the flux $h$ itself, can be measured. However, due to 
effects involving the creation and annihilation of virtual vortex-antivortex 
pairs which encircle the basepoint $x_0$ used to determine the flux $h$, 
this is only true if the scattering process is done sufficiently fast. 
Otherwise the $h$ vortex state will start to mix with conjugates. Indeed, if 
the process is done adiabatically, all one can measure in this manner is the 
conjugacy class of $h$ \ref\almp{M.~Alford, K.-M.~Lee, J.~March-Russell and 
J.~Preskill, Nucl. Phys. {\bf B384} (1992) 251.}. If the beam of particles 
starts off very far from the vortex, then the experiment necessarily takes a 
long time and we cannot distinguish conjugate vortices. Another way of saying 
this is that vortex states with specific flux labels are not asymptotic states 
in the theory. Only the conjugacy class label is well-defined asymptotically. 
All of this means that vortex ``types'' are labelled by the conjugacy classes 
in $H$, and that we may safely apply the concept of statistics to vortices 
within any such class. Unless explicitly stated otherwise, we will use the 
term indistinguishable in the weak sense in the remainder of this paper.

\subsec{Braid Groups and Vortex Statistics on $\IR^2$}

Consider then a system of $n$ vortices on the plane $\IR^2$ located at the
positions $x_1,\dots ,x_n$, and whose respective fluxes $h_1,\dots ,h_n$ 
(defined using the loops in \fig\fone{}) are all conjugate to one another. 
Denote the corresponding quantum ``flux eigenstate'' by $\ket{h_1,\dots ,h_n}$.
These states form a basis for the space ${\cal H}$ of all 
{\it indistinguishable} $n$-vortex states with fixed vortex positions 
$x_1,\dots ,x_n$. We may study the quantum statistics of these vortices by 
considering how the elements of ${\cal H}$ transform under all possible 
permutations of the vortices. More precisely, ${\cal H}$ will carry a 
representation of the $n$-string braid group of the plane, $B_n(\IR^2)$. This 
group may be generated by $n-1$ elements $\sigma_1,\sigma_2,\ldots ,
\sigma_{n-1}$, where $\sigma_i$ represents the local (counterclockwise)
exchange of the two adjacent vortices at $x_i$ and $x_{i+1}$. These obey 
several relations from which all others can be derived \ref\art{E.~Artin, Abh.
Math. Sem. Hamburg {\bf 4} (1926) 47.}:\foot{Throughout this paper we read 
{\it all} products --- whether they be of braids, elements of $H$, or loops 
used to define fluxes --- from right to left.}
\eqn\one{\eqalign{\sigma_i\sigma_{i+1}\sigma_i&=\sigma_{i+1}\sigma_i
\sigma_{i+1},      \,\,\, 1\leq i \leq n-2 , \cr \sigma_i\sigma_j&=\sigma_j
\sigma_i,     \,\,\, \left|i-j\right|\geq 2~.}}
$B_2(\IR^2)$ is generated by a single exchange and is isomorphic to the
additive group of the integers, while for $n\geq 3$, each $B_n(\IR^2)$ is a
torsion-free nonabelian group. The action of $B_n(\IR^2)$ on the flux 
eigenstates is determined from \wz
\eqn\none{\sigma_i\ket{h_1,\dots ,h_i,h_{i+1},\dots ,h_n}=\ket{h_1,\dots ,
h_ih_{i+1}h_i^{-1},h_i,\dots ,h_n}.}
The linear extension of this action to all of ${\cal H}$ defines a {\it 
permutation representation} \ref\djs{D.~J.~S.~Robinson, {\it A Course in the 
Theory of Groups} (Springer-Verlag, New York, 1982).} of $B_n(\IR^2)$. 
(Note that \none\ conserves the {\it total flux} $h_1h_2\cdots h_n$ of the
$n$-vortex system, which is defined as the path-ordered exponential of the 
gauge field around a loop, with basepoint $x_0$ at infinity, circling 
the entire set of vortices once in a counterclockwise manner. It also does not 
destroy the indistinguishability of the vortices.) This representation can be 
decomposed into its irreducible pieces, each of which defines a ``statistical 
superselection sector'' for the vortices (that is, an invariant subspace of 
${\cal H}$) \ref\amb{T.~D.~Imbo, C.~Shah~Imbo and E.~C.~G.~Sudarshan, Phys. 
Lett. B {\bf 234} (1990) 103.}.

\topinsert
\GraphIn{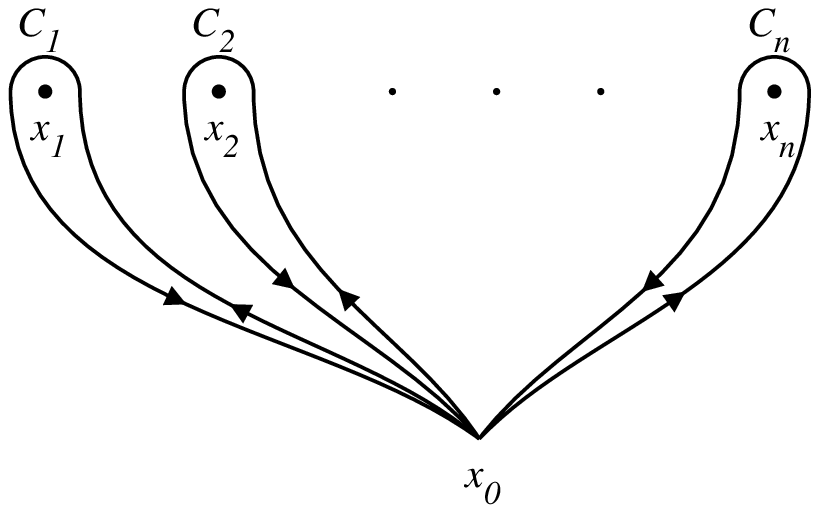}{3.33}{2.16}{1}{.5}{-3.56}
\noindent\baselineskip=12pt 
{\bf Figure 1.} A system of $n$ vortices on the plane.
The flux of the vortex located at $x_i$ is defined as the gauge field holonomy
around the loop $C_i$ which circles the vortex once in the counterclockwise
direction.
\endinsert

The irreducible unitary representations (IUR's) of $B_n(\IR^2)$, and hence the 
available statistical types for the vortices, can be determined with the help 
of the defining relations in \one . For example, we may consider the 
one-dimensional representations where each of the $\sigma_i$'s is represented 
by a simple phase. The first of the relations in \one\ (the {\it Yang Baxter 
relation}) requires that each such phase be the same (that is, $\sigma_i=
e^{{\rm i}\theta }$ independent of $i$), while the second relation is 
automatically satisfied in any one-dimensional representation. Note that there 
is no further restriction on $\theta$. The corresponding quantum statistics 
are called {\it fractional} (or $\theta$-) statistics, and the associated 
particles are {\it anyons} \ref\any{M.~Leinaas and J.~Myrheim, Nuovo Cimento 
B {\bf 37} (1977) 1\semi G.~A.~Goldin, R.~Menikoff and D.~H.~Sharp, J. Math. 
Phys. {\bf 22} (1981) 1664\semi F.~Wilczek, Phys. Rev. Lett. {\bf 48} (1982) 
144; {\bf 49} (1982) 957.}. The usual Bose and Fermi statistics correspond to 
$\theta =0$ and $\theta =\pi$, respectively. There are also higher-dimensional 
IUR's of $B_n(\IR^2)$ which yield {\it nonabelian anyons} 
\amb\ref\gms{G.~A.~Goldin, R.~Menikoff and D.~H.~Sharp, Phys. Rev. Lett. 
{\bf 54} (1985) 603.}\ref\non{G.~Moore and N.~Read, Nucl. Phys. {\bf B360} 
(1991) 362.}.

Since the gauge group $H$ is finite, there are only a finite number of flux 
eigenstates available (fixing the vortex positions) for any $n$-vortex system. 
In other words, ${\cal H}$ is finite-dimensional. Thus, only a finite quotient 
group of $B_n(\IR^2)$ acts nontrivially on the states in the above permutation 
representation, and the IUR's of this group then determine the allowed 
statistics of the indistinguishable vortices. (The specifics of the quotient 
group will depend on both $H$ and $n$.) One consequence of this fact is that 
the only types of fractional statistics that can be realized in a vortex 
system are those where the statistical angle $\theta$ is a rational multiple 
of $\pi$. Note also that a necessary and sufficient condition for the 
existence of {\it some} type of statistics other than Bose and Fermi is that 
$H$ is nonabelian. For examples of vortex systems on $\IR^2$ obeying 
fractional statistics, as well as other forms of exotic statistics, along with 
a discussion of the spin and $H$-charge properties of composite vortices, see 
\bdfi .

\newsec{Vortices on Open Surfaces --- Orientable and Nonorientable}
More generally, we may consider vortices moving on an open surface $M$ which
is not necessarily simply connected --- that is, $M$ may have noncontractible
loops. As examples we have the plane with an arbitrary number of punctures,
attached {\it handles}, and attached {\it cross-caps}.\foot{Attaching a handle 
to $\IR^2$, or indeed to any surface, can be viewed as first removing two open 
disks and then connecting the two resulting circular boundaries by a tube. 
Attaching a cross-cap to a surface can be thought of as first removing a single 
open disk and then identifying antipodal points on the resulting circular 
boundary.} On such a space, each noncontractible loop $\ell$ may also carry a 
flux. In other words, there exist configurations in which the path-ordered 
exponential of the gauge field around such a loop $\ell$ (based at $x_0$) is 
nontrivial.\foot{If $M$ has more than one open end, as in the case of the 
cylinder $\IR\times S^1$ (which is homeomorphic to the plane with a single 
puncture), then we choose one such end as ``infinity'' and locate our 
basepoint $x_0$ there.} The value of this flux will be invariant under 
continuous deformations of $\ell$, as long as these deformations do not cross 
a vortex. So a flux eigenstate is no longer simply labelled by the vortex 
fluxes, but also by the fluxes carried by the ``surface topology''.\foot{Note 
that under a global gauge transformation by $h\in H$, both the vortex {\it and}
surface topology fluxes get conjugated by $h$.} It of course suffices to 
specify the fluxes carried by a set of loops whose homotopy classes generate 
the fundamental group $\pi_1(M,x_0)$, since the flux of any other loop can be 
determined from these and the vortex fluxes. 

As on the plane, conjugate vortices are indistinguishable and we may study 
their statistics by considering the braid group representations carried by the 
$n$-vortex states. The new braid groups $B_n(M)$ still contain the exchanges
$\sigma_i$ described above, subject to the relations in \one . However, these
operators no longer suffice to generate the entire group. We also need a set
of generators which correspond to taking a single vortex (say that at $x_n$)
around a set of noncontractible loops in $M$ which generate $\pi_1(M,x_n)$.
Additional relations among these one-particle loops, as well as between these  
loops and the exchanges, are also needed to define the group. The elements of 
$B_n(M)$ then permute the flux eigenstates (in which both vortices and 
noncontractible loops carry flux) among themselves. This action respects the 
defining relations in $B_n(M)$, as well as keeps the total flux fixed. Here, 
the total flux is defined as the path-ordered exponential of the gauge field 
around the ``circle at infinity'' containing the basepoint $x_0$. One can 
again decompose the associated permutation representation into its irreducible 
pieces. In the superselection sector corresponding to a given IUR of $B_n(M)$, 
the statistics of the vortices is determined by how the subgroup generated by 
the $\sigma_i$'s is represented. (For a more precise treatment, see \amb .)

We can describe the above states and the corresponding braid group action in 
a slightly more mathematical language as follows. An $n$-vortex state on a 
general open surface $M$ may be viewed as a homomorphism $\psi:\pi_1(M_{(n)},
x_0)\to H$, where $M_{(n)}$ is the surface $M$ with $n$ punctures located at 
the positions of the vortices \buch . This map assigns an element of the 
unbroken gauge group $H$ to each vortex as well as to each noncontractible 
loop on $M$, while respecting the multiplication in $\pi_1(M_{(n)},x_0)$. Now, 
there is faithful representation of the braid group $B_n(M)$ as a group of 
automorphisms of $\pi_1(M_{(n)},x_0)$. That is, one can construct a one-to-one 
homomorphism $f:B_n(M)\to{\rm Aut}(\pi_1(M_{(n)},x_0))$. So to every braid 
$b\in B_n(M)$, there corresponds an isomorphism $b_*:\pi_1(M_{(n)},x_0)\to
\pi_1(M_{(n)},x_0)$. Each such $b$ acts on a quantum state 
$\psi:\pi_1(M_{(n)},x_0)\to H$ by precomposition with the inverse isomorphism 
$b_*^{-1}$ --- that is, $b$ sends $\psi$ to $\psi\circ b_*^{-1}$.

There is one further subtlety if $M$ is nonorientable --- for instance, in the 
examples described above when cross-caps are present. On these spaces, the 
orientation of the contour which enters the path-ordered exponential defining 
the flux of a vortex is not globally well-defined. Put another way, suppose 
that such an oriented contour is chosen around a given vortex in order to 
compute its flux $h$. Now take the vortex for an excursion around an 
orientation-reversing loop on the surface. When the vortex returns to its 
original position, a computation of its flux now yields (a conjugate of) 
$h^{-1}$. Using our previous definition of indistinguishability, this is 
enough to imply that inverse fluxes are indistinguishable on nonorientable 
surfaces. Of course, ordinary conjugate fluxes remain indistinguishable as 
well on these spaces. Thus, two vortices are indistinguishable on a 
nonorientable surface if and only if their fluxes are conjugate, or the 
flux of one is conjugate to the inverse flux of the other. In contrast to 
orientable surfaces, we see that distinct fluxes can be indistinguishable 
{\it even if $H$ is abelian}.

\newsec{Vortices on Compact Surfaces}

On an open surface, recall that we selected the basepoint at infinity. This
avoided problems associated with allowing vortices to circle the basepoint.
On a compact surface, however, there are no points at infinity at which to
locate the basepoint. We are thus forced to choose some arbitrary point $x_0$
as the basepoint, which by compactness cannot be ``too far away'' from the 
vortices under study. Hence, we are left with the problem that when any 
vortex circles the basepoint, all vortex fluxes become conjugated and the 
basepoint appears to be physical.  In other words, vortices can  ``scatter'' 
off the basepoint with seemingly observable consequences.

This problem can also be seen by examining the action of the braid group 
$B_n(M)$ on the $n$-vortex states on a compact surface $M$. Consider, for 
example, the case of three vortices on the sphere $S^2$. $B_3(S^2)$ can be 
presented as \ref\fbus{E.~Fadell and J.~Van Buskirk, Duke Math. Jour. {\bf 29} 
(1962) 243.}
$$ B_3(S^2)=\left\langle \sigma_1,\sigma_2 \left|
	\sigma_1\sigma_2\sigma_1=\sigma_2\sigma_1\sigma_2,\,
	\sigma_1\sigma_2^2\sigma_1=e \right.\right\rangle ,
$$
where $e$ is the identity element. The product $\sigma_1\sigma_2^2\sigma_1$
appearing in the second relation corresponds to taking the vortex at $x_1$ 
around the other two vortices. This operation is trivial on $S^2$ because 
the path of this vortex can be deformed around the ``other side'' of the 
sphere and contracted to a point. Now let $\ket{h_1,h_2,h_3}$ represent a 
state with vortex fluxes $h_1$, $h_2$, and $h_3$. A contour (based at $x_0$) 
which goes around all three vortices can be shrunk to a point in a similar
manner. Thus, the total flux $h_1h_2h_3$ of a three-vortex flux eigenstate on 
$S^2$ must be trivial. Also, the action of $B_3(S^2)$ on the states is 
generated by (see \none )
$$ \eqalign{
\sigma_1\ket{h_1,h_2,h_3} &= \ket{h_1h_2h_1^{-1},h_1,h_3} \cr
\sigma_2\ket{h_1,h_2,h_3} &= \ket{h_1,h_2h_3h_2^{-1},h_2}. \cr}
$$
The first of the braid group relations on $S^2$ is automatically satisfied, 
but the second is not as can be seen by calculating
$$ \eqalign{
\sigma_1\sigma_2^2\sigma_1\ket{h_1,h_2,h_3}
&= \sigma_1\sigma_2^2\ket{h_1h_2h_1^{-1},h_1,h_3}\cr
&= \sigma_1\sigma_2\ket{h_1h_2h_1^{-1},h_1h_3h_1^{-1},h_1}\cr
&= \sigma_1\ket{h_1h_2h_1^{-1},h_1h_3h_1h_3^{-1}h_1^{-1},h_1h_3h_1^{-1}}\cr
&= \ket{h_1h_2h_3h_1h_3^{-1}h_2^{-1}h_1^{-1},h_1h_2h_1^{-1},h_1h_3h_1^{-1}} \cr
&= \ket{h_1,h_1h_2h_1^{-1},h_1h_3h_1^{-1}},\cr}
$$
which is just $\ket{h_1,h_2,h_3}$ with all the fluxes conjugated by $h_1$.  
This can also be seen by noting that the path $\sigma_1\sigma_2^2\sigma_1$ 
is equivalent to taking the $h_1$ vortex once around the basepoint $x_0$, 
which we noted earlier has this effect. So we see that it is the basepoint 
that is getting in the way of deforming $\sigma_1\sigma_2^2\sigma_1$ to a 
point. Put another way, the vortices behave as if they are moving on the 
punctured surface $S^2 - \{ x_0\}$, which is not the true physical situation.

The solution to this problem is straightforward. To enforce the correct braid 
relations, the state $\ket{h_1,h_1h_2h_1^{-1},h_1h_3h_1^{-1}}$ must be 
identified with the original state $\ket{h_1,h_2,h_3}$. More generally, any 
two states which differ by an overall conjugation by some element $h\in H$ 
--- or in other words, a global gauge transformation by $h$ --- should be 
identified. This situation persists for any number of vortices on an arbitrary 
compact surface $M$. That is, $n$-vortex states on $M$ which differ by a 
global $H$ gauge transformation must be considered {\it identical}. This 
identification ensures that the basepoint is unphysical since encircling the 
basepoint causes an overall conjugation of fluxes, including those carried 
by the surface topology. It is also needed to yield a well-defined action of 
$B_n(M)$ on the states. The set of inequivalent flux eigenstates of a system 
of $n$ vortices on $M$ (with fixed vortex positions) may, as a result, be much 
smaller than one might have originally assumed, and likewise the corresponding 
finite quotient group of $B_n(M)$ which acts nontrivially on these states. A 
further consequence of this identification is that vortex systems on compact 
surfaces must have no net $H$ charge, although subsystems may still carry 
charge.\foot{We thank Hans Dykstra for bringing this to our attention.} Of 
course, all of the above points are moot if $H$ is abelian. These 
considerations are only nontrivial for systems of nonabelian vortices on 
compact surfaces.

From a more formal point of view, the source of the above phenomenon is that
there is no longer a natural homomorphism from $B_n(M)$ to ${\rm Aut}
(\pi_1(M_{(n)},x_0))$ when $M$ is compact. There is, however, a homomorphism 
from $B_n(M)$ to the {\it outer} automorphism group ${\rm Out}(\pi_1(M_{(n)},
x_0))$ --- that is, we must mod out ${\rm Aut}(\pi_1(M_{(n)},x_0))$ by the 
normal subgroup ${\rm Inn}(\pi_1(M_{(n)},x_0))$ of {\it inner} automorphisms 
(conjugations) of $\pi_1(M_{(n)},x_0)$, Out=Aut/Inn. As a result, flux 
eigenstates are no longer described by homomorphisms from $\pi_1(M_{(n)},x_0)$ 
to $H$, but by conjugacy classes of such maps so that the action of the braid 
group on these states is well-defined. We denote the image of the map 
$f:B_n(M)\to {\rm Out}(\pi_1(M_{(n)},x_0))$ by $V_n(M)$, and call it the {\it 
$n$-string vortex braid group of $M$}. $V_n(M)$ is the more relevant group 
for classifying the $n$-vortex states on the compact manifold $M$, and 
studying their statistics. If $M$ is orientable, then $V_n(M)$ is a subgroup 
of the $n$-th {\it mapping class group} of $M$, and is computed in 
\ref\bir{J.~Birman, Comm. Pure Appl. Math. {\bf 22} (1969) 213.}.

It is useful to note that the above flux eigenstates on a compact surface
will always satisfy a ``total flux constraint''. This generalizes what we have  
already encountered for three vortices on $S^2$, namely, $h_1h_2h_3=e$. To 
see this, recall that every such surface $M$ (without boundary) is 
homeomorphic to a sphere with a number of attached handles and cross-caps
\ref\mas{W.~S.~Massey, {\it A Basic Course in Algebraic Topology} 
(Springer-Verlag, New York 1991).}. This attached surface topology can, up to 
homeomorphism, be taken as localized on the sphere. In a flux eigenstate 
describing $n$ vortices on $M$, the flux computed around a contour $C$ (based 
at $x_0$) enclosing all of the vortices as well as all of this surface 
topology must be trivial since $C$ can be contracted to a point around the 
other side of the sphere. (That is, $C$ represents the identity element in 
$\pi_1(M_{(n)},x_0)$.) This relation among all of the vortex fluxes and
the fluxes of the noncontractible loops is what we call the total flux 
constraint. Similarly, the generators of the braid group $B_n(M)$ will satisfy 
an analog of the ``compactness relation'' $\sigma_1\sigma_2^2\sigma_1=e$ for 
three vortices on $S^2$. This will read $b=e$, where $b$ is a braid taking one 
of the vortices for an excursion around a loop in $M$ enclosing all of the 
other vortices and all of the surface topology. It is this relation which is 
not satisfied unless we identify states differing by a global $H$ gauge 
transformation.

One might ask how a system of vortices on a compact surface $M$ with the 
states ``modded out'' by global gauge transformations as above is related to 
the corresponding system on the punctured surface $M_*=M-\{ x_0\}$. After all, 
if the vortices and the surface topology are localized in a ``small'' region 
of $M$ far from $x_0$, shouldn't the physics (and hence the number of states) 
be the same as if the vortices were on the open surface $M_*$ (where conjugate 
states are distinct)?\foot{We thank Kai-Ming Lee for discussions on this 
point.} The difference between the two cases is in the presence or absence of 
points at ``spatial infinity''. In the open surface case, one can imagine 
introducing any number of additional (unphysical) vortices at infinity with 
{\it fixed} fluxes to use as a reference when trying to determine the flux of 
a given vortex. For example, these fluxes might be part of a detector which is 
far away from the interaction region in a scattering experiment. All vortex 
fluxes can then be measured relative to these fixed fluxes at infinity, and 
states which differ by an overall conjugation of the original {\it physical} 
(that is, not at infinity) fluxes, including those of the surface topology, 
are distinct. Note that the reference fluxes themselves should not be allowed 
to be conjugated --- they must remain fixed to provide an absolute reference. 
Similarly, unphysical $H$-charges can also be introduced at infinity to 
provide a reference for Aharonov-Bohm scattering. The addition of the reference 
fluxes or charges at infinity does not change the braid group (or its action 
on the states) since they are never really in the space.

When space is compact, however, one is not free to introduce additional 
reference fluxes or charges, because there are no points at infinity where 
they can be ``hidden''.  Any such particles which one introduces become 
physical, and the braid group and its action change --- that is, the particles
in your detector must be taken into account in the braid group just like any 
other particles. The freedom on a compact surface to perform a global $H$ gauge 
transformation at any fixed time without changing the state then allows one to 
fix the flux of {\it any} given vortex throughout its history.\foot{This is not 
quite true on {\it nonorientable} compact surfaces, where we can only fix the 
flux of a vortex up to inversion.} (That is, one can fix a ``global gauge''.) 
Depending on the other fluxes present and the particulars of the group $H$, 
there may still be some global gauge freedom left which can then be used to 
fix the fluxes of an additional subset of vortices. The fluxes of the remaining 
vortices can now be measured relative to this chosen set of vortices. In 
particular, states which differ by an overall conjugation of the fluxes of the 
``unchosen'' vortices and the surface topology (but not the fluxes of the 
chosen, fixed vortices) are indeed distinct. Thus, the chosen vortices on the 
compact surface serve the same function as the ``detector'' or reference 
vortices in the open surface case. All of the results of the open case can be 
recovered from the compact case by expanding the surface and pulling the 
chosen, fixed vortices off to infinity (along with the basepoint $x_0$) where 
they become unphysical reference vortices. The states of the remaining, 
physical vortices are not modded out by global gauge transformations. 
Equivalently, if a scattering experiment is performed in a small region of a 
compact surface, then the reference vortices used in the detector can be taken 
as the chosen vortices and the results are the same as if the experiment had 
been done on an open surface using only the unchosen vortices. Similar 
considerations apply if the detector contains $H$-charges instead of fluxes. 
In either case, modding out by global gauge transformations is equivalent to 
fixing the global gauge in the detector. 

\newsec{Examples}
\subsec{Vortices on $S^2$}
For any $n\geq 1$, the braid group $B_n(S^2)$ \fbus\ has the ``same'' set of 
generators $\sigma_i$, $i=1,\dots ,n-1$, as for $B_n(\IR^2)$. However, along
with \one , we have the additional compactness relation
$$\sigma_1\cdots\sigma_{n-2}\sigma_{n-1}^2\sigma_{n-2}\cdots\sigma_1=e.$$ 
This additional constraint (which generalizes the $n=3$ relation discussed 
earlier) is enough to make the braid group finite for $n=2$ and $n=3$. 
More specifically, $B_2(S^2)=\IZ_2$ and $B_3(S^2)$ is a nonabelian group of 
order 12. However, $B_n(S^2)$ remains infinite and nonabelian (although no 
longer torsion-free) for each $n\geq 4$. This relation also restricts the 
possible statistical behavior of indistinguishable particles on the sphere 
\ref\sph{D.~J.~Thouless and Y.-S.~Wu, Phys. Rev. B {\bf 31} (1985) 1191\semi 
J.~S.~Dowker, J. Phys. A {\bf 18} (1985) 3521.}. For example, in a 
one-dimensional IUR of $B_n(S^2)$ all the $\sigma_i$'s are equal to a single 
phase $e^{{\rm i}\theta}$, which by the compactness relation must be a 
$(2n-2)$-th root of unity. Thus, unlike on the plane, only a subset of the 
full range of fractional statistics is allowed. Similar results hold for the 
higher-dimensional IUR's of $B_n(S^2)$ and the corresponding {\it nonabelian} 
statistics \amb\ref\ahp{E.~C.~G.~Sudarshan, T.~D.~Imbo and C.~Shah~Imbo, Ann. 
Inst. Henri Poincar\'e {\bf 49} (1988) 387.}. 

When we consider indistinguishable {\it vortices} on $S^2$, further
restrictions arise. The $n$-vortex flux eigenstate $\ket{h_1,h_2,\dots ,h_n}$ 
on $S^2$ must satisfy the total flux constraint $h_1h_2\cdots h_n=e$, and must 
be considered {\it identical} to any state which can be obtained from it by a
global $H$ gauge transformation (see above) --- that is, to any state
$\ket{h_1^{\prime} ,\dots ,h_n^{\prime}}$ where $h_i^{\prime}=hh_ih^{-1}$ for
some $h\in H$ and all $i=1,\dots , n$. As a result, certain elements of 
$B_n(S^2)$ always act trivially on the set of inequivalent quantum states for 
{\it any} unbroken gauge group $H$. (As on $\IR^2$, the action of the 
$\sigma_i$'s is given by \none .) To see this, lets look at the nontrivial 
braid $\Delta\equiv(\sigma_1\cdots\sigma_{n-1})^n$ which squares to the 
identity element in $B_n(S^2)$. $\Delta$ rotates the entire $n$-vortex system 
by 2$\pi$, and has the same effect on a state as a global gauge transformation 
by the total flux $h_1h_2\cdots h_n$. Thus, $\Delta$ acts trivially on every 
state. (Note that although $h_1h_2\cdots h_n$ is trivial here, we did not 
need to use this.) Since $\Delta$ has trivial action so does any conjugate of 
$\Delta$ in $B_n(S^2)$, and the quotient group of $B_n(S^2)$ obtained by 
adding the single relation $\Delta=e$ to those above is the vortex braid group 
$V_n(S^2)$ discussed earlier. That is, $V_n(S^2)$ is the quotient of $B_n(S^2)$
by the normal closure of the subgroup generated by $\Delta$. This is the 
relevant group for classifying the quantum states and their statistics.

This result has an immediate consequence for the possible fractional 
statistics of $n$ indistinguishable vortices on $S^2$. Namely, when the 
compactness relation is combined with $\Delta=e$, we see that for $n$ odd 
the only allowed statistical phases are $(n-1)$-th roots of unity. That 
is, in the one-dimensional IUR's of $V_n(S^2)$, only these phases are allowed 
for the $\sigma_i$'s. For example, if $n=3$ this means that only Bose and 
Fermi statistics are possible. The ``semions'' ($\theta =\pi /2$) allowed by 
$B_3(S^2)$ are now ruled out. For even $n$, we do not obtain any information 
on fractional statistics that we didn't already know from $B_n(S^2)$. Of 
course some choices of $H$ will restrict the possibilities further. However, 
as far as we can tell, there is some choice of $H$ (and indistinguishable 
vortex states satisfying the total flux constraint) which yields any desired 
type of statistics allowed by $V_n(S^2)$, with one exception. This is the 
case $n$=2 which we now discuss.

Consider a state $\ket{h_1,h_2}$ consisting of two indistinguishable 
vortices on the sphere. The fluxes $h_1$ and $h_2$ must be conjugate by 
indistinguishability; let's say $h_2=hh_1h^{-1}$. Also, the total flux of 
the two vortices must be trivial so that $h_2=h_1^{-1}$.\foot{As a curiosity, 
note that if $H$ has the property that no nontrivial element is conjugate to 
its inverse, then there is no configuration consisting solely of two 
indistinguishable vortices with nontrivial fluxes on the sphere.} In 
particular, $h_1$ and $h_2$ necessarily commute. Thus, the exchange $\sigma$ 
simply interchanges the two fluxes; $\sigma\ket{h_1,h_2}=\ket{h_2,h_1}$. 
However, a global gauge transformation by $h^{-1}$ shows that this is 
equivalent to the original state. Thus $\sigma$ acts trivially on the states, 
and only Bose statistics is possible. By contrast, the group $V_2(S^2)$ is 
generated by $\sigma$ and is isomorphic to $\IZ_2$. But because of the power of 
the total flux constraint in a two-vortex system on $S^2$, we see that the 
fermionic representation of $V_2(S^2)$ cannot be realized.

\subsec{Vortices on $\IR P^2$}
Another simple compact surface is the projective plane $\IR P^2$ , which can be 
viewed as a sphere with antipodal points identified. This identification 
introduces noncontractible loops into the space --- namely those which, in 
this picture, start at a given point $x_0$ and end at the antipodal point of 
$x_0$. (A frame dragged along such a loop will reverse its orientation, 
showing that $\IR P^2$ is nonorientable.) The square of any such loop is 
homotopically trivial, so that the fundamental group $\pi_1(\IR P^2,x_0)$ is 
isomorphic to $\IZ_2$. The projective plane is also homeomorphic to the sphere
with one attached cross-cap. A noncontractible loop in this picture starts at
some point $x_0$ on the sphere and eventually hits a point $y$ on the boundary 
of the removed open disk. It then emerges from the antipodal point of $y$ 
before returning to $x_0$. Due to the above, there will be an additional 
generator in the braid groups of $\IR P^2$. More precisely, $B_n(\IR P^2)$ is 
generated by the usual exchanges $\sigma_i$, $i=1,\dots ,n$, as well as an 
element $\rho$ which corresponds to taking one of the particles (say that at 
$x_n$) around the noncontractible loop on $\IR P^2$ based at $x_n$ (see 
\fig\ftwo{}). The defining relations between these generators are quite 
cumbersome in general. A presentation of $B_2(\IR P^2)$, which is a nonabelian 
group of order 16, is
\eqn\lbr{ \left\langle 
\sigma, \rho \left| (\sigma\rho)^{-2}(\rho\sigma)^2=\sigma^2,\, 
\rho^2\sigma^2=e \right. \right\rangle .}
For $n\geq 3$, $B_n(\IR P^2)$ is infinite and nonabelian. Presentations can be
found in \ref\bus{J.~Van Buskirk, Trans. Amer. Math. Soc. {\bf 122} (1966) 
81.}. For any $n$ it can be shown that fractional statistics are not possible 
for indistinguishable particles on $\IR P^2$ \ref\mri{T.~D.~Imbo and 
J.~March-Russell, Phys. Lett. B {\bf 252} (1990) 84.}. That is to say, in any 
IUR $D$ of $B_n(\IR P^2)$ where each $\sigma_i$ is represented by 
$e^{{\rm i}\theta }I$, $I$ being the identity matrix, we must have $\theta =0$ 
or $\pi$. (Indeed, the dimension of $D$ must be 1.) Hence, $D$ can only yield 
Bose or Fermi statistics for the particles.

\topinsert
\GraphIn{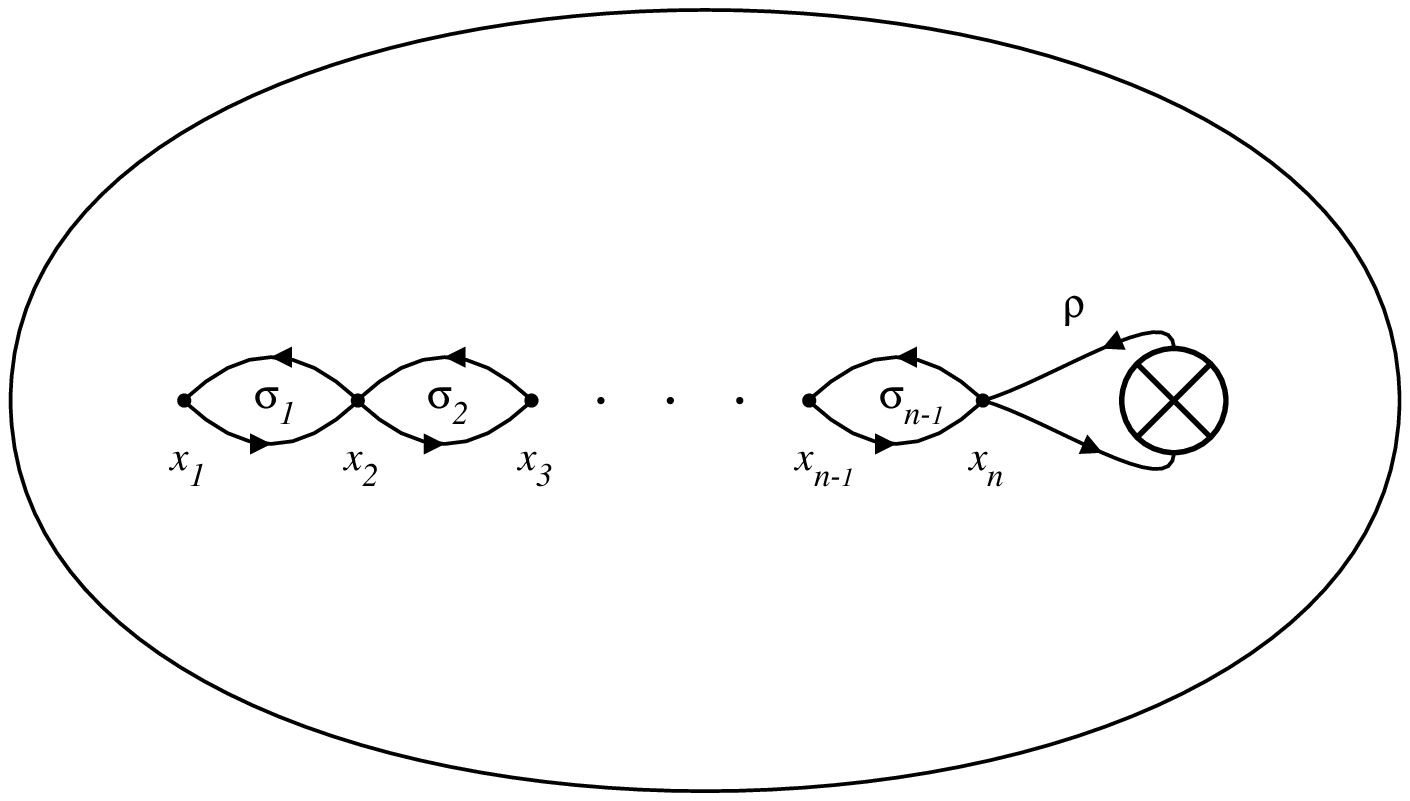}{5.62}{2.79}{1}{-.22}{-4.26}
\noindent\baselineskip=12pt 
{\bf Figure 2.} The generating operations of $B_n(\IR P^2)$, the braid group 
of $n$ particles on the projective plane viewed as a sphere with an attached
cross-cap. Here $\sigma _i$ is the local counterclockwise exchange of the
vortices at $x_i$ and $x_{i+1}$, while $\rho $ takes the vortex located at 
$x_n$ around a noncontractible loop in $\IR P^2$, shown as a path going once 
through the cross-cap.
\endinsert

\topinsert
\GraphIn{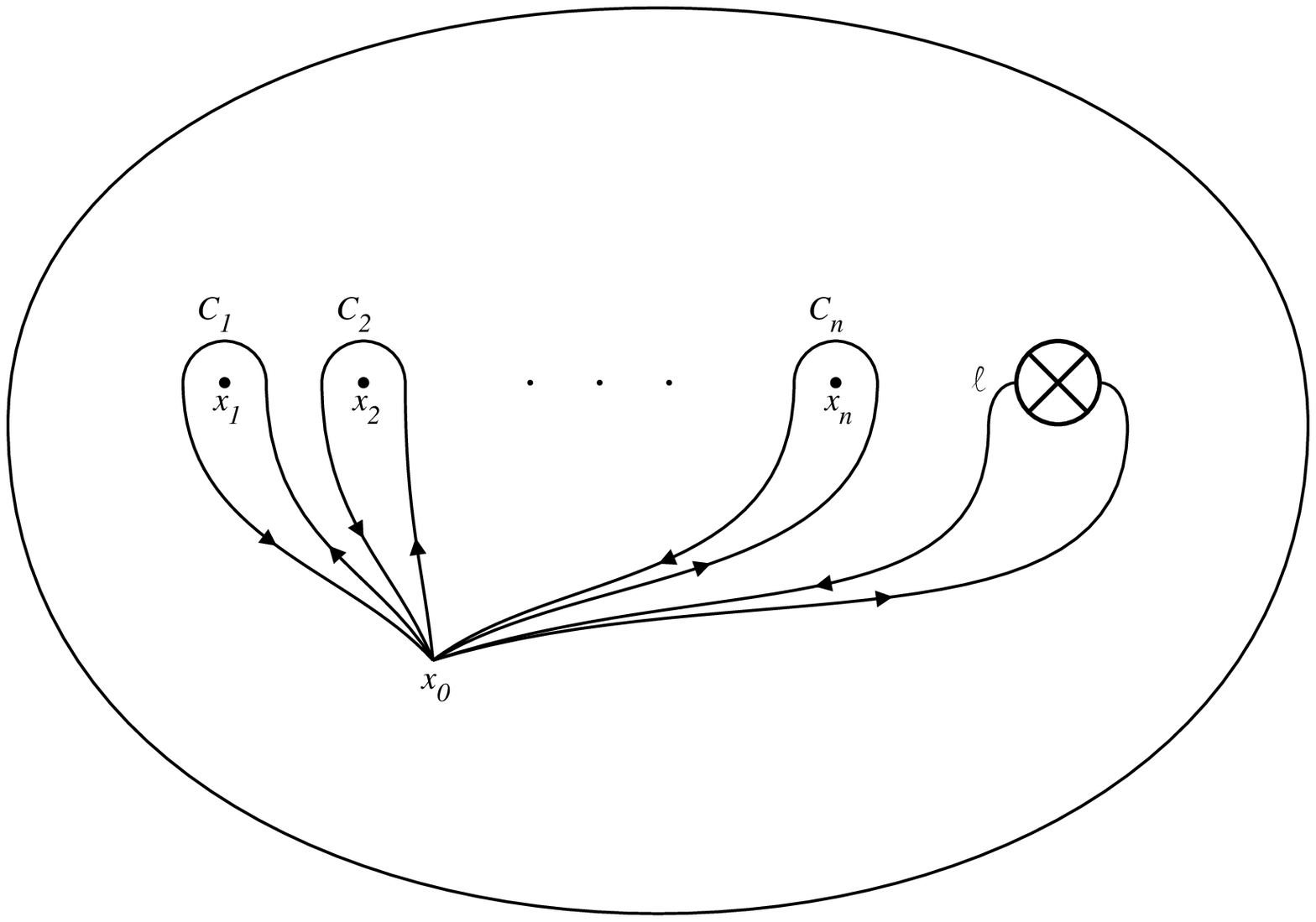}{5.62}{4.07}{1}{-1.00}{-4.70}
\noindent\baselineskip=12pt 
{\bf Figure 3.} A system of $n$ vortices on $\IR P^2$, viewed as a sphere with
an attached cross-cap. The flux of the vortex located at $x_i$ is defined as
the gauge field holonomy around the loop $C_i$ which circles the vortex once in
the counterclockwise direction. A noncontractible loop $\ell $ in $\IR P^2$ can
also carry a flux, defined as the gauge field holonomy around $\ell $. Here
$\ell $ is depicted as a loop going once through the cross-cap. The homotopy 
classes of the $C_i$'s and $\ell$ generate $\pi_1(\IR P^2_{(n)},x_0)$.
\endinsert

Recall that in a gauge theory with finite unbroken gauge group $H$, the 
path-ordered exponential of the gauge field around a noncontractible loop can 
give nontrivial results. That is, such loops can carry a nontrivial flux in 
$H$. Thus, a flux eigenstate describing $n$ vortices on $\IR P^2$ is labeled 
by the individual vortex fluxes $h_i$ (measured using loops $C_i$ based at 
$x_0$) as well as the flux $r$ of the noncontractible loop $\ell$ based at 
$x_0$ (see \fig\fthree{}). We denote such a state by $\ket{h_1,\dots ,h_n;r}$. 
In the absence of vortices, we must have $r^2=e$ since the square of $\ell$ 
can be shrunk to a point. However, for $n\geq 1$ we see that $\ell^2$ is 
homotopic to a loop which circles the entire set of vortices once in a 
clockwise fashion. Therefore, the total flux constraint on the above 
$n$-vortex state reads $h_1h_2\cdots h_nr^2=e$. Also, if the vortices are 
indistinguishable, then for each pair of fluxes $h_i$ and $h_j$ in the above 
state, $h_i$ is conjugate to either $h_j$ or $h_j^{-1}$ (since $\IR P^2$ is 
nonorientable). The action of $B_n(\IR P^2)$ on the states 
$\ket{h_1,\dots ,h_n;r}$ is generated by
$$ \eqalign{
\sigma_i\ket{h_1,\dots ,h_i,h_{i+1},\dots h_n;\,r}&=\ket{h_1,\dots ,
h_ih_{i+1}h_i^{-1},h_i,\dots ,h_n;r} \cr
\rho\ket{h_1,\dots ,h_n;\,r}&=\ket{h_1,\dots ,h_{n-1},
(h_nr)h_n^{-1}(h_nr)^{-1};h_nr} .\cr}
$$
Again, not all elements of $B_n(\IR P^2)$ have nontrivial action on the states. 
Due to the total flux constraint and modding out by global gauge 
transformations, the element $\Delta\equiv(\sigma_1\cdots\sigma_{n-1})^n\in 
B_n(\IR P^2)$ once again acts trivially on every state, although this time the 
reasoning is slightly more subtle. As for $S^2$, $\Delta$ is still homotopic 
to a 2$\pi$-rotation of the $n$ vortices (within a disk), having the effect of 
conjugating each vortex flux $h_i$ by $h_1h_2\cdots h_n$, the total flux of 
the $n$-vortex subsystem. However, the flux $r$ of the noncontractible loop is 
unchanged. Now $h_1h_2\cdots h_n$ is not trivial in general, but instead equal 
to $r^{-2}$ by the total flux constraint. Hence, a global gauge transformation 
by $r^2$ will restore the vortex fluxes to their original values, and still
leave $r$ unchanged (since $r^2$ commutes with $r$). Thus, $\Delta$ and its 
conjugates in $B_n(\IR P^2)$ act trivially, and the states of the system 
decompose into IUR's of the vortex braid group $V_n(\IR P^2)$ --- the group 
obtained from $B_n(\IR P^2)$ by adjoining the single relation $\Delta =e$. We 
believe that this is the whole story --- that is, each IUR of $V_n(\IR P^2)$ 
can be realized with some $H$.

As an example consider the case $n=2$. As advertised, the action of $\Delta=
\sigma^2$ on the states is easily seen to be trivial from \lbr . When we add 
the relation $\sigma^2=e$ to $B_2(\IR P^2)$, we obtain $V_2(\IR P^2)$ which
is isomorphic to the dihedral group of order 8. In contrast to the case of two 
particles on a sphere, the above group allows Fermi statistics (as well as
Bose, of course). The smallest choice of the unbroken gauge group $H$ for which
this occurs is the unique nonabelian group of order 21, which can be presented
as
$$\left\langle s,t \left| t^3=e, t^{-1}st=s^2 \right. \right\rangle~.$$
Since $s$ and $s^2$ are conjugate in $H$, and $s^7=e$, the state
$$ \eqalign{
\ket{\Psi}&= \ket{s,s^2;s^4}-\ket{s,s^4;s}-\ket{s,s^3;s^5}+\ket{s^3,s;s^5} \cr
	&+ \ket{s^3,s^6;s^6}-\ket{s^3,s^5;s^3}-\ket{s^3,s^2;s}+\ket{s,s^5;s^4}
		\cr }
$$
describes two indistinguishable vortices on $\IR P^2$. $\ket{\Psi}$ is an 
eigenstate of both $\sigma$ and $\rho$, with eigenvalues $-1$ and $+1$
respectively. Thus, this state defines a one-dimensional IUR of $V_2(\IR P^2)$ 
in which the vortices obey Fermi statistics. Note that in order to obtain this 
result we must use the fact that conjugate states are identical.

If the two vortices are instead moving on the open surface $\IR P^2_*=\IR P^2
-\{ x_0\}$, which is homeomorphic to the plane with a single attached 
cross-cap, then there are several differences from the above analysis. First,
the total flux constraint $h_1h_2r^2=e$ on a state $\ket{h_1,h_2;r}$ is no
longer required. Second, conjugate states are no longer considered identical 
in general. And finally the braid group $B_2(\IR P^2_*)$, although still 
generated by the ``same'' elements $\sigma$ and $\rho$ which act on the states 
$\ket{h_1,h_2;r}$ as above, no longer contains the compactness relation 
$\sigma^2\rho^2=e$. That is, we can present $B_2(\IR P^2_*)$ as
$$ \left\langle 
\sigma, \rho \left| (\sigma\rho)^{-2}(\rho\sigma)^2=\sigma^2\right. 
\right\rangle .
$$
Of course conjugate and inverse conjugate vortices remain indistinguishable.
Also, even though the infinite, nonabelian group $B_2(\IR P^2_*)$ is much 
larger than $V_2(\IR P^2)$, it is straightforward to show that fractional 
statistics still cannot occur.

\subsec{Vortices on $T^2$}
The final compact surface that we will consider is the torus $T^2=S^1\times 
S^1$, which is homeomorhic to the sphere with a single attached handle. The 
braid group $B_n(T^2)$ can be generated by the $\sigma_i$'s, as well as single 
particle excursions $\alpha$ and $\beta$ around the two noncontractible loops 
shown in \fig\ffour{}. Again, the defining relations are somewhat complicated.
For example, the infinite, nonabelian group $B_2(T^2)$ can be presented as
$$ \eqalign{
\langle \sigma, \alpha, \beta &| 
\sigma^2\alpha\beta\alpha^{-1}\beta^{-1}=e,
(\sigma\alpha)^2=(\alpha\sigma)^2, \cr
&\,(\sigma\beta^{-1})^2=(\beta^{-1}\sigma)^2,
\beta\sigma\alpha\sigma=\sigma\alpha\sigma^{-1}\beta
\rangle  .\cr}
$$
For presentations when $n\geq 3$, see \ref\tor{J.~Birman, Comm. Pure Appl. Math.
{\bf 22} (1969) 41\semi Y. Ladegaillerie, Bull. Sci. Math. {\bf 100} (1976) 255.}. 
For any $n$, one can show that the only statistics associated with the 
one-dimensional IUR's of $B_n(T^2)$ are Bose and Fermi \amb\ahp . However, 
some types of fractional statistics can be obtained by considering 
higher-dimensional IUR's \mri\ref\ein{T.~Einarsson, Phys. Rev. Lett. {\bf 64} 
(1990) 1995.}. That is, there are IUR's of dimension greater than one in which 
each of the $\sigma_i$'s is represented by the scalar matrix 
$e^{{\rm i}\theta }I$. However, the allowed values of $\theta$ depend on $n$ 
and are all rational multiples of $\pi$.

\topinsert
\GraphIn{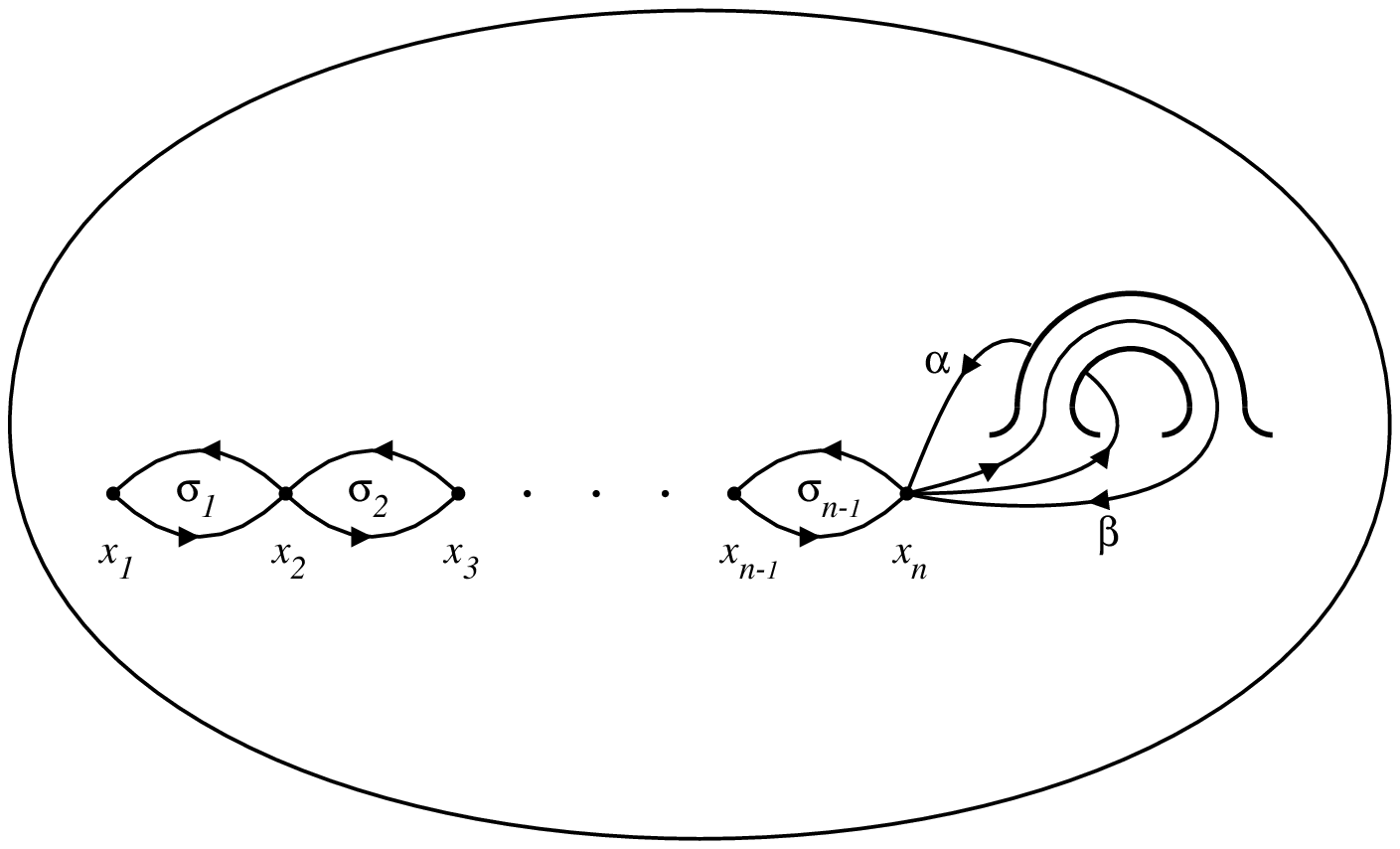}{5.62}{3.00}{1}{-1.64}{-4.69}
\noindent\baselineskip=12pt 
{\bf Figure 4.} The generating operations of $B_n(T^2)$, the braid group of
$n$ particles on the torus viewed as a sphere with an attached handle. Again
$\sigma _i$ is the local counterclockwise exchange of the vortices at $x_i$ and
$x_{i+1}$. The operations $\alpha $ and $\beta $ take the vortex located at $x_n$
around the noncontractible loops in $T^2$ shown as paths through and along the
handle respectively.
\endinsert

\topinsert
\GraphIn{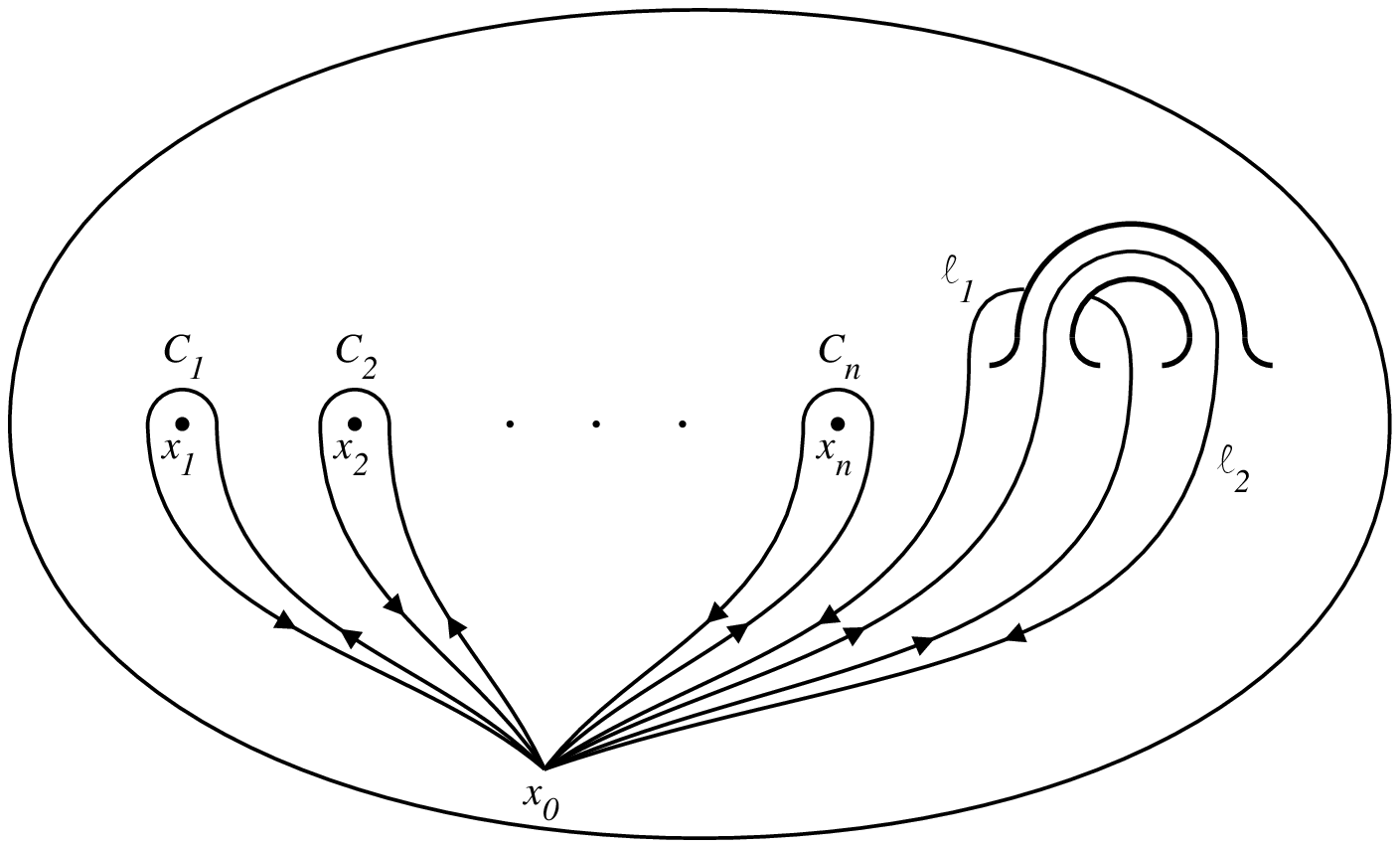}{5.69}{3.00}{1}{-1.61}{-4.41}
\noindent\baselineskip=12pt 
{\bf Figure 5.} A system of $n$ vortices on $T^2$ viewed as a sphere
with an attached handle. The flux of the vortex located at $x_i$ is defined as
the gauge field holonomy around the loop $C_i$ which circles the vortex once in
the counterclockwise direction. The flux carried by any noncontractible loop in
$T^2$ can be generated from the gauge field holonomies around the loops
$\ell _1$ and $\ell _2$ which pass once through and along the handle. The 
homotopy classes of the $C_i$'s and $\ell_i$'s generate $\pi_1(T^2_{(n)},x_0)$.
\endinsert

Let us now move to the spontaneously broken gauge theories under consideration.
The fundamental group $\pi_1(T^2,x_0)$ is isomorphic to $\IZ\times\IZ$, and is
generated by (the homotopy classes of) the two loops $\ell_1$ and $\ell_2$ 
shown in \fig\ffive{}. Each of these loops can carry a nontrivial flux in $H$. 
Call these fluxes $a$ and $b$ respectively. If there are no vortices on the 
surface, then $a$ and $b$ must commute in $H$ since the homotopy classes of 
$\ell_1$ and $\ell_2$ commute in $\pi_1(T^2,x_0)$. However, if there are $n$ 
vortices on the surface (with fluxes $h_i$ measured using the loops $C_i$ in 
\ffive , $i=1,\dots ,n$), then the ``commutator'' $\ell_1\ell_2\ell_1^{-1}
\ell_2^{-1}$ is homotopic to a loop which circles the entire set of vortices 
once in a clockwise fashion. Therefore, the general total flux constraint reads 
$h_1\cdots h_naba^{-1}b^{-1}=e$. We write the corresponding flux eigenstate as 
$\ket{h_1,\dots ,h_n;a,b}$. $B_n(T^2)$ acts on these states via (see also 
\kml )
\eqn\ntwo{\eqalign{
\sigma_i\ket{h_1,\dots ,h_n;a,b}&=\ket{h_1,\dots ,
h_ih_{i+1}h_i^{-1},h_i,\dots ,h_n;a,b} \cr
\alpha\ket{h_1,\dots ,h_n;a,b}&=\ket{h_1,\dots ,h_{n-1},(h_na)h_n(h_na)^{-1};
h_nah_n^{-1},bh_n^{-1}} \cr
\beta\ket{h_1,\dots ,h_n;a,b}&=\ket{h_1,\dots ,h_{n-1},bh_nb^{-1};
bh_n^{-1}b^{-1}h_nabh_nb^{-1},(bh_n^{-1})b(bh_n^{-1})^{-1}} .\cr}}
One may have guessed that the element $\Delta$ again has trivial action on all 
states. But this is no longer true since the flux $h_1h_2\cdots h_n$ of the 
$n$-vortex subsystem does not commute with $a$ and $b$ in general, although 
it commutes with their commutator. Hence, after performing a 2$\pi$-rotation 
of the $n$-vortices (within a disk), there is in general no global gauge 
transformation which will give us back all the original fluxes.

\topinsert
\GraphIn{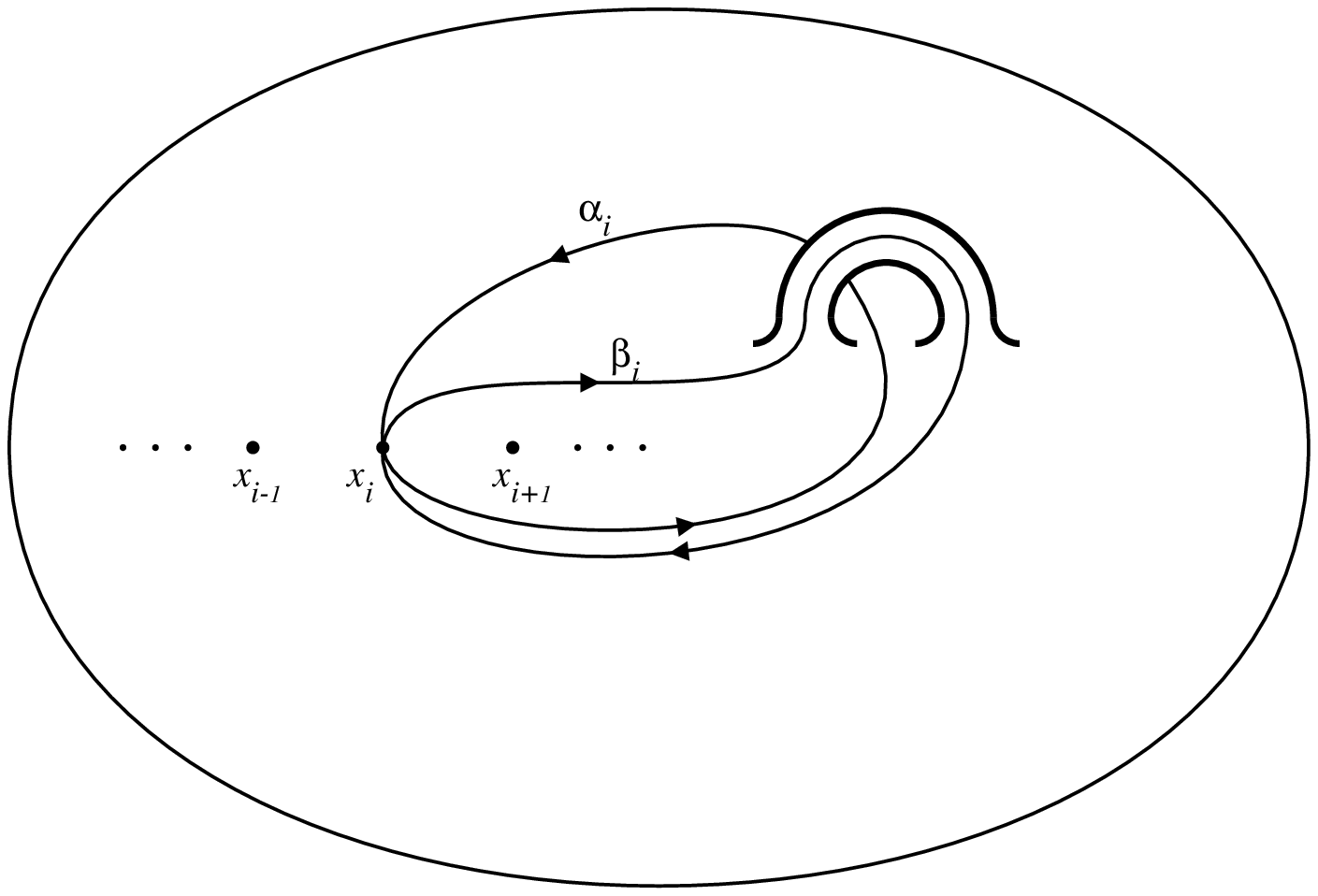}{5.62}{3.33}{1}{-.28}{-4.59}
\noindent\baselineskip=12pt 
{\bf Figure 6.} The operation $\alpha _i$ takes the
vortex located at $x_i$ around a noncontractible loop passing once through the
attached handle, and also circling the vortices located at $x_j$ for $j>i$ once
in the counterclockwise direction. The operation $\beta _i$ takes the vortex
located at $x_i$ around a noncontractible loop passing once along the attached
handle, and also circling the vortices located at $x_j$ for $j>i$ once in the
clockwise direction. 
\endinsert

Alternatively, consider the braids $\alpha_i\equiv\sigma_i\cdots\sigma_{n-1} 
\alpha\sigma_{n-1}\cdots\sigma_i$, $i=1,\dots n$. $\alpha_i$ takes the vortex 
at $x_i$ around the loop shown in \fig\fsix{}. (Note that $\alpha_n=\alpha$, 
and that $\alpha_i$ and $\alpha_j$ commute for all $i$ and $j$.) From \ntwo\ 
it is straightforward to show that the product $A\equiv\alpha_1\cdots
\alpha_n$ conjugates all the fluxes in a state $\ket{h_1,\dots ,h_n;a,b}$ by 
$a$, and thus gives us back the original state. Similarly, if we consider the 
braids (see \fsix ) $\beta_i\equiv\sigma_i^{-1}\cdots\sigma_{n-1}^{-1}\beta
\sigma_{n-1}^{-1}\cdots\sigma_i^{-1}$, $i=1,\dots n$, then the product 
$B\equiv\beta_1\cdots\beta_n$ conjugates all the fluxes in a state 
$\ket{h_1,\dots ,h_n;a,b}$ by $b$. Hence, the normal closure of the subgroup 
of $B_n(T^2)$ generated by $A$ and $B$ acts trivially on all the quantum 
states. It can further be shown that these are the only such elements --- that 
is, the quotient of $B_n(T^2)$ by this normal subgroup (or equivalently, 
$B_n(T^2)$ along with the additional relations $A=B=e$) yields the vortex braid 
group $V_n(T^2)$ \bir . Although these new relations do restrict the 
representations of $B_n(T^2)$ that can be realized in a vortex system, the 
existence of Bose, Fermi and fractional statistics (with statistical angle a 
rational multiple of $\pi$) are unaffected.

When $n=1$ we have $A=\alpha$ and $B=\beta$, and these elements generate the
entire braid group $B_1(T^2)=\pi_1(T^2)=\IZ\times\IZ$. Thus, the vortex braid
group $V_1(T^2)$ is trivial. This means that {\it any} one-vortex state 
$\ket{h;a,b}$ on the torus is unchanged after the vortex traverses a loop 
on $T^2$. When $n=2$, the elements $A$ and $B$ are just $(\sigma\alpha)^2$ 
and $(\sigma^{-1}\beta)^2$ respectively. Using the presentation of $B_2(T^2)$
given above, one obtains 
$$ V_2(T^2) = \left\langle \sigma, \alpha, \beta \left| 
(\sigma\alpha)^2=e,
(\beta^{-1}\sigma)^2=e,
(\beta^{-1}\sigma\alpha)^2=e
\right.\right\rangle ,
$$
which is isomorphic to the free product $\IZ_2*\IZ_2*\IZ_2$. This group allows
the existence of Bose, Fermi and semionic statistics for the two vortices. 
For example, if we take $H$ to be the 16-element group given by
$$ \left\langle s,t \left| t^2=e, tst=s^3 \right. \right\rangle ,
$$
then the state 
$$ \eqalign{
\ket{\Psi}&=\ket{s^2,s^6;s^1,s^1}-\ket{s^2,s^6;s^1,s^3}
		+\ket{s^2,s^6;s^1,s^5}-\ket{s^2,s^6;s^1,s^7} \cr
	&+\ket{s^2,s^6;s^3,s^1}-\ket{s^2,s^6;s^3,s^3}
		+\ket{s^2,s^6;s^3,s^5}-\ket{s^2,s^6;s^3,s^7} \cr
	&+\ket{s^2,s^6;s^5,s^1}-\ket{s^2,s^6;s^5,s^3}
		+\ket{s^2,s^6;s^5,s^5}-\ket{s^2,s^6;s^5,s^7} \cr
	&+\ket{s^2,s^6;s^7,s^1}-\ket{s^2,s^6;s^7,s^3}
		+\ket{s^2,s^6;s^7,s^5}-\ket{s^2,s^6;s^7,s^7} \cr}
$$
is an eigenstate of $\sigma$, $\alpha$, and $\beta$, with eigenvalues $-1$,
$-1$, and $+1$ respectively. Thus, the vortices are fermions in this state. 
(Note that $s^2$ and $s^6$ are conjugate in $H$ so that the two vortices are 
indistinguishable, and that $s^8=e$ so that the total flux constraint is met.
Also, the identification of conjugate states must be used to obtain our 
result.) While we fully expect that there exists some choice of finite gauge 
group $H$ which leads to semionic two-vortex states on $T^2$, we have found 
that such statistics do not manifest themselves using any of the groups of 
order 21 or less (as well as many other classes of larger finite groups).

For comparison, lets consider the case of two vortices on the open surface
$T^2_*=T^2-\{ x_0\}$ --- that is, the torus with a single puncture. This space
can also be thought of (up to homeomorphism) as the plane with a single handle 
attached. The total flux constraint $h_1h_2aba^{-1}b^{-1}$ on a state 
$\ket{h_1,h_2;a,b}$ is no longer needed on $T^2_*$, and conjugate states are 
not considered identical in general. The braid group $B_2(T^2_*)$ has the 
``same'' generators as $B_2(T^2)$, acting on the states in the same way. 
However, the defining relations are different. More precisely, $B_2(T^2_*)$ 
can be presented as
$$ 
\langle \sigma, \alpha, \beta | 
(\sigma\alpha)^2=(\alpha\sigma)^2,
(\sigma\beta^{-1})^2=(\beta^{-1}\sigma)^2,
\beta\sigma\alpha\sigma=\sigma\alpha\sigma^{-1}\beta
\rangle ,
$$
which differs from $B_2(T^2)$ by the deletion of the compactness relation
$\sigma^2\alpha\beta\alpha^{-1}\beta^{-1}=e$. This group allows for the 
existence of anyonic two-vortex states with any statistical angle $\theta$
that is a rational multiple of $\pi$, while we have seen that only the values 
$\theta =0,\pi /2,\pi ,3\pi /2$ are allowed on $T^2$.

\newsec{Conclusions}

We have studied the physics of topological vortices moving on an arbitrary 
two-dimensional manifold in a spontaneously broken gauge theory where a
simply connected gauge group $G$ breaks to a finite subgroup $H$. Our work
contains four central points:

\vskip 12pt
\noindent
(1) On an arbitrary surface, vortices with conjugate fluxes can be considered 
to be indistinguishable. For vortices on a nonorientable surface, ``inverse
conjugate'' fluxes are also indistinguishable.

\vskip 12pt
\noindent
(2) For systems of $n$ indistinguishable vortices on an open surface $M$, 
there is a ``faithful'' action of the braid group $B_n(M)$ on the quantum 
states. Here, by faithful we mean that there are no elements of $B_n(M)$
which act trivially on all states {\it for all choices of $H$}.

\vskip 12pt
\noindent
(3) For vortices on a compact surface $M$, two states which differ by a
global $H$ gauge transformation must be considered {\it identical}.\foot{An 
alternative way of viewing this result using Gauss' law is given in 
\ref\lolee{K.-M.~Lee and H.-K.~Lo, hep-th/9505108.}. (The e-print \lolee\  
appeared after one of its authors was informed of our work.) Other related 
items are also discussed here.} As a result, there exists a normal subgroup of 
$B_n(M)$ which acts trivially on all states, and we denote by $V_n(M)$ the 
corresponding quotient group of $B_n(M)$. This ``vortex braid group'' $V_n(M)$ 
then acts faithfully on the states in the sense described above (except in the 
case of two vortices on $S^2$).

\vskip 12pt
\noindent
(4) Each of the above results has consequences for the quantum statistics
of vortex systems, which we illustrated in a number of examples.

\vskip 12pt
The results in this paper can be extended in several directions. For example,
instead of restricting our attention to systems containing only ``pure 
fluxes'', we can more generally consider particle-like objects carrying both
flux and $H$ charge. This requires a generalization of our formalism which has 
strong connections to the theory of quantum groups (in particular, the {\it 
quantum double} $D(H)$ of the finite group $H$) \lp\kml\ref\bdp{F.~A.~Bais, 
P.~van~Driel and M.~de~Wild~Propitius, Phys. Lett. {\bf B280} (1992) 63\semi
M.~de~Wild~Propitius and F.~A.~Bais, hep-th/9511201.}. We have also implicitly 
assumed in our work that the Lagrangian of the theory does not contain a 
Chern-Simons term for the $G$ gauge field. The inclusion of such a term 
requires yet another extension of the formalism \ref\bdpt{F.~A.~Bais, 
P.~van~Driel and M.~de~Wild~Propitius, Nucl. Phys. {\bf B393} (1993) 547.}.

Finally, when $M$ is compact and orientable, we can ask the following question: 
Why has only the subgroup $V_n(M)$ of the $n$-th mapping class group 
${\cal M}_n(M)$ played a role in our analysis? Or, in other words, is 
there a physical interpretation of the ``remainder'' of ${\cal M}_n(M)$? The 
answer is yes. To obtain the full mapping class group we must add to $V_n(M)$ 
additional generators corresponding to motions of the handles on $M$. That is, 
up until now we have treated the geometry of $M$ as a fixed background. But if
the metric on $M$ is elevated to a dynamical variable in our theory, then the 
$n$-vortex states on $M$ are classified according to representations of ${\cal 
M}_n(M)$. The fluxes carried by these dynamical handles (or {\it geons} \ref
\geo{J.~L.~Friedman and R.~D.~Sorkin, Phys. Rev. Lett. {\bf 44} (1980) 1100; 
{\bf 45} (1980) 148 (E)\semi C.~Aneziris, A.~P.~Balachandran, M.~Bourdeau, 
S.~Jo, T.~R.~Ramadas and R.~D.~Sorkin, Int. J. Mod. Phys. {\bf A4} (1989) 5459.}) 
will not only change when vortices pass through them as above, but also, in
general, as the handles pass around and through each other. (Indeed, even when 
there are no vortices present the zero-th mapping class group ${\cal M}_0(M)$ 
is nontrivial and acts on the quantum states corresponding to flux-carrying 
handles alone.) In this context we may speak of the indistinguishability of 
handles and discuss their quantum statistics. Similar results also apply to 
handles on open manifolds, and to cross-caps on nonorientable spaces. We will 
address these and related issues in more detail in a forthcoming paper.

\bigskip

\noindent
\centerline{{\bf Acknowledgements}}
\vskip 12pt
\noindent
It is a pleasure to thank Hans Dykstra and Kai-Ming Lee for many useful 
discussions. This work was supported in part by the U.~S.~Department of 
Energy under contract number DE-FG02-91ER40676.

\listrefs
\bye